\documentclass[english,journal=cmatex,manuscript=article]{achemso}
\usepackage[T1]{fontenc}
\usepackage[latin9]{inputenc}
\usepackage{color}
\usepackage{array}
\usepackage{booktabs}
\usepackage{textcomp}
\usepackage{multirow}
\usepackage{amsmath}
\usepackage{amsthm}
\usepackage{graphicx}

\makeatletter

\title{Comment on ``Structure Prediction of Li\textminus Sn and Li\textminus Sb
Intermetallics for Lithium-Ion Batteries Anodes''}
\author{Raja Sen}
\author{Priya Johari}
\email{priya.johari@snu.edu.in, psony11@gmail.com}
\affiliation{Department of Physics, School of Natural Sciences, Shiv Nadar University,
Greater Noida, Gautam Buddha Nagar, UP 201 314, India.}
\providecommand{\tabularnewline}{\\}

\numberwithin{equation}{section}
\numberwithin{figure}{section}

\usepackage{lmodern}

\makeatother

\usepackage{babel}
\begin{document}
\captionsetup[figure]{name=Figure} 
\renewcommand\thefigure{\arabic{figure}}
\textcolor{black}{\newpage}

\textcolor{black}{In a recently published article Mayo et al.\cite{Martin@2017}
presented the ground state crystal structures of various experimentally
unknown Li-Sn intermetallic compounds at ambient pressure (\textasciitilde{}
0 GPa) and 0 K temperature using ab-initio random structure searching
method (AIRSS) with high-throughput screening from the Inorganic Crystal
Structure Database (ICSD).\cite{AIRSS} In their study,\cite{Martin@2017}
besides the experimentally known phases of Li-Sn such as, $\mathrm{Li_{2}Sn_{5}}$
($\mathrm{P4/mbm}$),\cite{Robert_2007,Dunlap_1999,Li2Sn5} $\mathrm{Li_{1}Sn_{1}}$
($\mathrm{P2/m}$, $\mathrm{I4_{1}/amd}$),\cite{Robert_2007,Dunlap_1999,alpha_LiSn,beta_LiSn}
$\mathrm{Li_{7}Sn_{3}}$ ($\mathrm{P2_{1}/m}$),\cite{Robert_2007,Dunlap_1999,Li7Sn3}
$\mathrm{Li_{5}Sn_{2}}$ ($\mathrm{R\bar{3}m}$),\cite{Robert_2007,Dunlap_1999,Li5Sn2}
$\mathrm{Li_{13}Sn_{5}}$ ($\mathrm{P\bar{3}m1}$),\cite{Robert_2007,Dunlap_1999,Li13Sn5}
$\mathrm{Li_{7}Sn_{2}}$ ($\mathrm{Cmmm}$),\cite{Robert_2007,Dunlap_1999,Li7Sn2}
and $\mathrm{Li_{17}Sn_{4}}$ ($\mathrm{F\bar{4}3m}$),\cite{Robert_2007,Dunlap_1999,Li17Sn4_Gowar,Li17Sn4_Lipu}
Mayo et al. also reported two previously unknown stable phases for
Li-Sn such as, $\mathrm{Li_{8}Sn_{3}}$-$\mathrm{R\bar{3}m}$ and
$\mathrm{Li_{7}Sn_{2}}$-$\mathrm{P\bar{1}}$ along with several Li-Sn
metastable phases ($\mathrm{Li_{1}Sn_{2}}$, $\mathrm{Li_{2}Sn_{3}}$,
$\mathrm{Li_{7}Sn_{9}}$, $\mathrm{Li_{3}Sn_{2}}$, $\mathrm{Li_{5}Sn_{3}}$,
$\mathrm{Li_{2}Sn}_{1}$, $\mathrm{Li_{3}Sn_{1}}$, $\mathrm{Li_{4}Sn_{1}}$,
$\mathrm{Li_{5}Sn_{1}}$, and $\mathrm{Li_{7}Sn_{1}}$) which lie
within 20 meV/atom from the convex hull tie-line. However, while going
through their article, we noticed a significant inconsistency and
contradictions in their results. Moreover, a one-to-one comparison
with our published results\cite{Sen@2017} revealed a disagreement
in the symmetry of $\mathrm{Li_{3}Sn_{1}}$, $\mathrm{Li_{7}Sn_{2}}$,
$\mathrm{Li_{4}Sn_{1}}$, $\mathrm{Li_{5}Sn_{1}}$, and $\mathrm{Li_{7}Sn_{1}}$
discussed by Mayo et al.\cite{Martin@2017} }

\textcolor{black}{Predicting the lowest energy ground state structures
is always an important task, as that is what determines the properties
of any material and specifically, in the context of Li-ion batteries
this is particularly crucial, since the open circuit voltage, Li-diffusivity,
and swelling upon lithiation, etc., depend critically on the crystal
structure and its respective ground state energy. We, therefore, re-calculated
the formation energy of the reported phases (given as CIF files in
the ``Supporting Information'' by Mayo et al.\cite{Martin@2017})
and our predicted phases, by considering higher and strict convergence
criteria (see the ``Methodology and }Computational\textcolor{black}{{}
Details in SI'') and compared the results on an equal footing by
calculating the formation enthalpy of each Li-Sn compound. Interestingly,
current calculations reveal that Mayo et al.\cite{Martin@2017} not
only predicted completely wrong structures in few cases (much higher
in energy than what we found, in some cases a difference of more than
100 meV/atom), but for many other cases it is not just that they published
a wrong symmetry for the right structure but the structures themselves
are often wrong, which can mislead the scientific community. Through
this comment, we therefore not only highlight several disparities
in reporting the correct crystal structure and describing the proper
symmetry of several Li-Sn compounds by Mayo et al.\cite{Martin@2017}
but, also present the correct ground state structures of Li-Sn compounds
at ambient pressure (\textasciitilde{} 0 GPa) and 0 K temperature,
which are extremely important to know, in order to establish an in-depth
understanding of the charge-discharge process in Li-Sn batteries. }

\section*{LOWEST ENERGY Li-Sn STRUCTURES}

\begin{figure}[t]
\begin{centering}
fig2\includegraphics[scale=0.13]{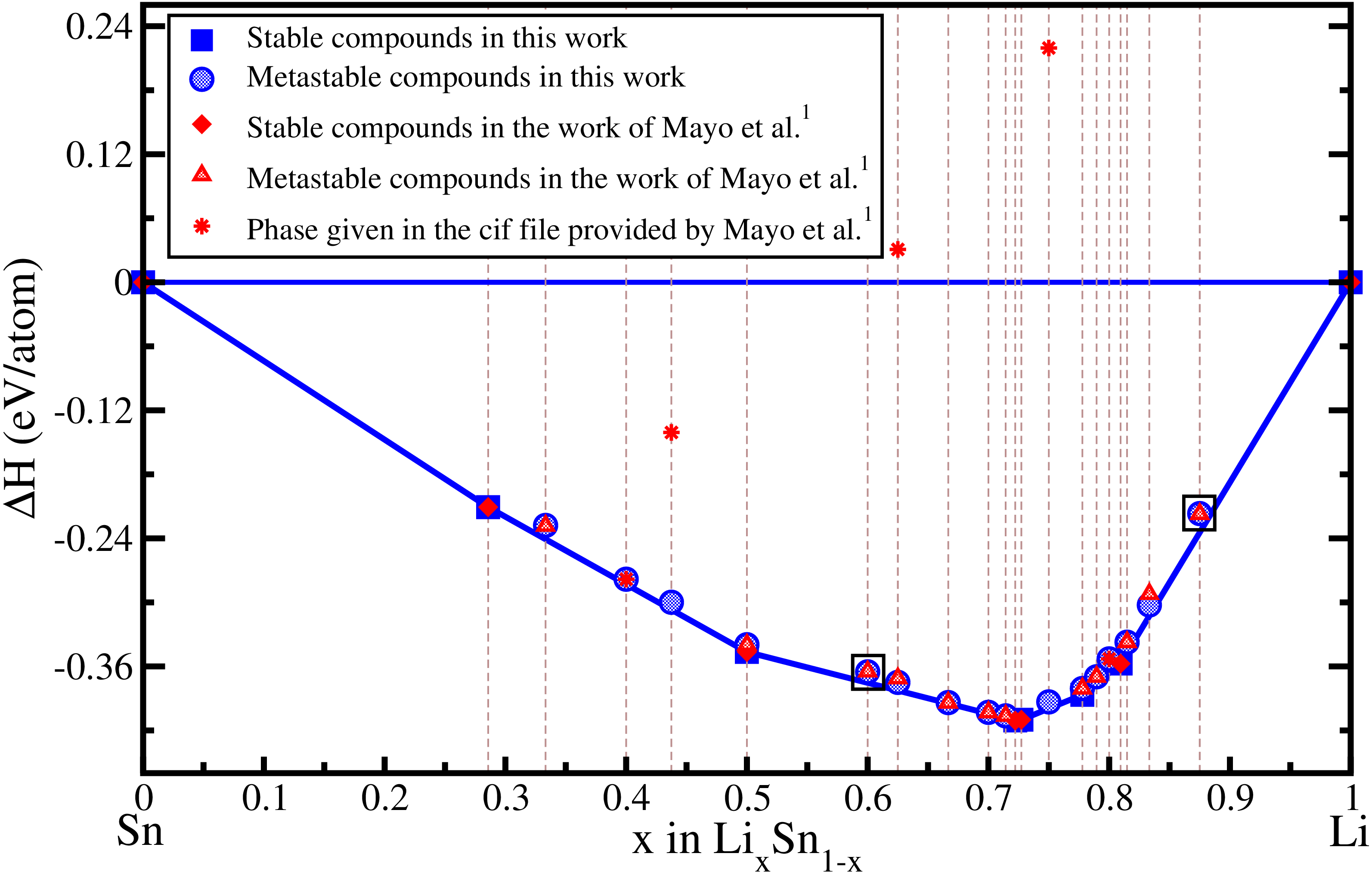}
\par\end{centering}
\centering{}\caption{Convex hull for the Li-Sn system at ambient pressure (\textasciitilde{}
0 GPa) and 0 K temperature. For $\mathrm{Li_{7}Sn_{9}}$ (x = 0.4375),
$\mathrm{Li_{5}Sn_{3}}$ (x = 0.625), $\mathrm{Li_{3}Sn_{1}}$ (x
= 0.75) and $\mathrm{Li_{5}Sn_{1}}$ (x = 0.833), current calculations
predict the correct ground state structures and phase, whose formation
enthalpies are lower than the respective phase of the compounds reported
by Mayo et al.\textcolor{black}{\cite{Martin@2017}} Here, red colored
stars represent the compounds for which wrong CIF files are provided
\textcolor{black}{by Mayo et al.\cite{Martin@2017} in their ``Supporting
Information'', while} the points enclosed by rectangular box represent
the stoichiometries ($\mathrm{Li_{3}Sn_{2}}$ and $\mathrm{Li_{7}Sn_{1}}$)
for which our calculations have predicted different phase with comparable
formation energy with respect to the phase of the respective compound,
reported by Mayo et al.\textcolor{black}{\cite{Martin@2017}} \label{1atm_chull}}
\end{figure}

\textcolor{black}{In search of accurate ground state structures of
the stable and metastable Li-Sn compounds at ambient pressure and
0 K temperature, current calculations following strict convergence
criterion reveal few new phases whose formation enthalpies are either
comparable or lower than the respective phase of the compounds reported
by Mayo et al.\cite{Martin@2017} Also, on comparing our results with
those of Mayo et al.,\cite{Martin@2017} it has been noticed that
in most of the cases, either they have wrongly identified the ground
state structure (symmetry) or there have been involved contradictions
in presenting the proper space group and corresponding crystal structure
of the compounds in their article (Figure \ref{1atm_chull} and Table
\ref{convex-hull-table}). In order to present a concise comparison
and inconsistencies within the results reported by Mayo et al.,\cite{Martin@2017}
we have tabulated the description of crystal symmetry of stable and
metastable Li-Sn compounds in Table \ref{convex-hull-table}. The
table displays the crystal symmetry reported by Mayo et al. in the
``ABSTRACT'', ``Table 1'', ``RESULTS'', and ``Supporting Information''
of their article\cite{Martin@2017}, together with the symmetry and
their respective fitness (vertical distance from the convex hull tie
line) obtained by present calculations for each composition. The contradicting
results (within their paper and with respect to current work using
better convergence criteria) are highlighted in red colored fonts
in Table \ref{convex-hull-table}. The CIF files corresponding to
the newly predicted structures from current work have also been supplied
in the Supporting Information of this comment, while correction for
the symmetry and structure of contradicting  Li-Sn compounds (marked
in red in Table \ref{convex-hull-table}) are given next.}

\begin{table}
\begin{raggedright}
\caption{A \textcolor{black}{one-to-one co}mparison between the results presented
by Mayo et al.\cite{Martin@2017} and this work (in which calculations
are performed with higher convergence criterion). This table also
highlights the inconsistency (in red fonts) in presenting the correct
ground state symmetry and incorrect prediction of strcutures of several
Li-Sn compounds, in the recent work of Mayo et al.\cite{Martin@2017}
at ambient pressure and 0 K temperature. \label{convex-hull-table}}
\par\end{raggedright}
\begin{raggedright}
{\tiny{}}%
\begin{tabular}{cccccc}
\toprule 
\multirow{3}{*}{{\tiny{}Stoichiometry}} & \multicolumn{3}{c}{{\tiny{}Results of Mayo et al.}\textcolor{black}{\cite{Martin@2017}}} &  & {\tiny{}Results of this work}\tabularnewline
\cmidrule{2-6} 
 & \multicolumn{4}{c}{{\tiny{}Crystal symmetry described for ambient pressure in }} & {\tiny{}Crystal symmetry described }\tabularnewline
\cmidrule{2-5} 
 & {\tiny{}ABSTRACT} & {\tiny{}Table 1 $^{\#\#}$} & {\tiny{}RESULTS \& DISCUSSION} & {\tiny{}Supporting Information} & {\tiny{}for ambient pressure}\tabularnewline
\midrule
\midrule 
{\tiny{}Sn} & - & {\tiny{}$\mathrm{I4_{1}/amd}$ (0)} & - & - & {\tiny{}$\mathrm{I4_{1}/amd}$ (0)}\tabularnewline
\midrule 
{\tiny{}Li$_{2}$Sn$_{5}$$^{*}$} & - & {\tiny{}$\mathrm{P4/mbm}$ (0)} & - & - & {\tiny{}$\mathrm{P4/mbm}$(0)}\tabularnewline
\midrule 
{\tiny{}Li$_{1}$Sn$_{2}$} & {\tiny{}$\mathrm{P4/mmm}$} & {\tiny{}$\mathrm{P4/mmm}$ (8)} & {\tiny{}$\mathrm{P4/mmm}$} & {\tiny{}$\mathrm{P4/mmm}$} & {\tiny{}$\mathrm{P4/mmm}$ (10)}\tabularnewline
\midrule 
\textcolor{red}{\tiny{}Li$_{2}$Sn$_{3}$} & \textcolor{red}{\tiny{}$\mathrm{P\overline{1}}$} & \textcolor{red}{\tiny{}$\mathrm{P\overline{1}}$ (6)} & \textcolor{red}{\tiny{}$\mathrm{P\overline{1}}$} & \textcolor{red}{\tiny{}$\mathrm{P4/mmm}$ } & \textcolor{red}{\tiny{}$\mathrm{P4/mmm}$ (5)}\tabularnewline
\midrule 
\textcolor{red}{\tiny{}Li$_{7}$Sn$_{9}$} & \textcolor{red}{\tiny{}$\mathrm{P4_{2}/n}$ } & \textcolor{red}{\tiny{}$\mathrm{P4_{2}/n}$ (19)} & \textcolor{red}{\tiny{}$\mathrm{P4_{2}/}n$} & \textcolor{red}{\tiny{}$\mathrm{C222}$} & \textcolor{red}{\tiny{}$\mathrm{P2/m}$ (7)}\tabularnewline
\midrule 
\multirow{2}{*}{{\tiny{}Li$_{1}$Sn$_{1}$$^{*}$}} & \multirow{2}{*}{-} & {\tiny{}$\mathrm{P2/m}$ (0)} & \multirow{2}{*}{-} & \multirow{2}{*}{-} & {\tiny{}$\mathrm{P2/m}$ (0)}\tabularnewline
 &  & {\tiny{}$\mathrm{I4_{1}/amd}$ (-)} &  &  & {\tiny{}$\mathrm{I4_{1}/amd}$ (7)}\tabularnewline
\midrule 
\textcolor{red}{\tiny{}Li$_{3}$Sn$_{2}$} & \textcolor{red}{\tiny{}$\mathrm{P2_{1}/m}$} & \textcolor{red}{\tiny{}$\mathrm{P2_{1}/m}$ (12)} & \textcolor{red}{\tiny{}$\mathrm{P2_{1}/m}$} & \textcolor{red}{\tiny{}$\mathrm{P2_{1}/m}$} & \textcolor{red}{\tiny{}$\mathrm{Cmcm}$ (10)}\tabularnewline
\midrule 
\multirow{2}{*}{\textcolor{red}{\tiny{}Li$_{5}$Sn$_{3}$}} & \multirow{2}{*}{\textcolor{red}{\tiny{}$\mathrm{Im\overline{3}m}$}} & \multirow{2}{*}{\textcolor{red}{\tiny{}$\mathrm{Im\overline{3}m}$ (5)}} & \multirow{2}{*}{\textcolor{red}{\tiny{}$\mathrm{P\overline{4}3m}$ \& $\mathrm{Im\overline{3}m}$ }} & \textcolor{red}{\tiny{}$\mathrm{P\overline{4}3m}$} & \textcolor{red}{\tiny{}$\mathrm{R32}$ (7)}\tabularnewline
 &  &  &  & \textcolor{red}{\tiny{}(Not thermodynamically stable)} & \textcolor{red}{\tiny{}$\mathrm{Fmm2}$ (7)}\tabularnewline
 &  &  &  &  & \textcolor{red}{\tiny{}$\mathrm{Im\overline{3}m}$ (11)}\tabularnewline
\midrule 
{\tiny{}Li$_{2}$Sn$_{1}$} & {\tiny{}$\mathrm{Cmcm}$} & {\tiny{}$\mathrm{Cmcm}$ (1)} & {\tiny{}$\mathrm{Cmcm}$} & {\tiny{}$\mathrm{Cmcm}$} & {\tiny{}$\mathrm{Cmcm}$ (1)}\tabularnewline
\midrule 
{\tiny{}Li$_{7}$Sn$_{3}$} & - & {\tiny{}$\mathrm{P2_{1}/m}$ (2)} & - & - & {\tiny{}$\mathrm{P2_{1}/m}$ (1)}\tabularnewline
\midrule 
{\tiny{}Li$_{5}$Sn$_{2}$} & - & {\tiny{}$\mathrm{R\overline{3}m}$ (2)} & - & - & {\tiny{}$\mathrm{R\overline{3}m}$ (2)}\tabularnewline
\midrule 
{\tiny{}Li$_{13}$Sn$_{5}$$^{*}$} & - & {\tiny{}$\mathrm{P\overline{3}m1}$ (0)} & - & - & {\tiny{}$\mathrm{P\overline{3}m1}$ (0)}\tabularnewline
\midrule 
{\tiny{}Li$_{8}$Sn$_{3}$$^{*}$} & {\tiny{}$\mathrm{R\overline{3}m}$} & {\tiny{}$\mathrm{R\overline{3}m}$ (0)} & {\tiny{}$\mathrm{R\overline{3}m}$} & {\tiny{}$\mathrm{R\overline{3}m}$} & {\tiny{}$\mathrm{R\overline{3}m}$ (0)}\tabularnewline
\midrule 
\multirow{2}{*}{\textcolor{red}{\tiny{}Li$_{3}$Sn$_{1}$}} & \multirow{2}{*}{\textcolor{red}{\tiny{}$\mathrm{P3_{2}}$}} & \multirow{2}{*}{\textcolor{red}{\tiny{}$\mathrm{P3_{2}}$ (6)}} & \multirow{2}{*}{\textcolor{red}{\tiny{}(Not discussed)}} & \textcolor{red}{\tiny{}$\mathrm{P3m1}$} & \multirow{2}{*}{\textcolor{red}{\tiny{}$\mathrm{P2/m}$ (7)}}\tabularnewline
 &  &  &  & \textcolor{red}{\tiny{}(Not thermodynamically stable)} & \tabularnewline
\midrule 
\multirow{2}{*}{\textcolor{red}{\tiny{}Li$_{7}$Sn$_{2}$}} & \multirow{2}{*}{\textcolor{red}{\tiny{}$\mathrm{P\overline{1}}$}} & \textcolor{red}{\tiny{}$\mathrm{P\overline{1}}$ (0)} & \multirow{2}{*}{\textcolor{red}{\tiny{}$\mathrm{P\overline{1}}$}} & \multirow{2}{*}{\textcolor{red}{\tiny{}$\mathrm{P\overline{3}m1}$ $^{\#}$}} & \textcolor{red}{\tiny{}$\mathrm{P\overline{3}m1}$ (0)}\tabularnewline
 &  & \textcolor{red}{\tiny{}$\mathrm{Cmmm}$ (6)} &  &  & \textcolor{red}{\tiny{}$\mathrm{Cmmm}$ (6)}\tabularnewline
\midrule 
{\tiny{}Li$_{15}$Sn$_{4}$} & - & {\tiny{}$\mathrm{I\overline{4}3d}$ (6)} & - & - & {\tiny{}$\mathrm{I\overline{4}3d}$ (6)}\tabularnewline
\midrule 
\textcolor{red}{\tiny{}Li$_{4}$Sn$_{1}$} & \textcolor{red}{\tiny{}$\mathrm{P2_{1}}$} & \textcolor{red}{\tiny{}$\mathrm{P2_{1}}$ (13)} & \textcolor{red}{\tiny{}Not discussed} & \textcolor{red}{\tiny{}$\mathrm{R\overline{3}m}$} & \textcolor{red}{\tiny{}$\mathrm{R\overline{3}m}$ (13)}\tabularnewline
\midrule 
{\tiny{}Li$_{17}$Sn$_{4}$$^{*}$} & - & {\tiny{}$\mathrm{F\overline{4}3m}$ (0)} & - & - & {\tiny{}$\mathrm{F\overline{4}3m}$ (0)}\tabularnewline
\midrule 
{\tiny{}Li$_{22}$Sn$_{5}$} & - & {\tiny{}$\mathrm{F\overline{4}3m}$ (11)} & - & - & {\tiny{}$\mathrm{F\overline{4}3m}$ (10)}\tabularnewline
\midrule 
\textcolor{red}{\tiny{}Li$_{5}$Sn$_{1}$} & \textcolor{red}{\tiny{}$\mathrm{P6/mmm}$} & \textcolor{red}{\tiny{}$\mathrm{Pmma}$ (19)} & \textcolor{red}{\tiny{}Not discussed} & \textcolor{red}{\tiny{}$\mathrm{P6/mmm}$} & \textcolor{red}{\tiny{}$\mathrm{C2/m}$ (11)}\tabularnewline
\midrule 
\textcolor{red}{\tiny{}Li$_{7}$Sn$_{1}$} & \textcolor{red}{\tiny{}$\mathrm{C2}$} & \textcolor{red}{\tiny{}$\mathrm{Fmmm}$ (18)} & \textcolor{red}{\tiny{}Not discussed} & \textcolor{red}{\tiny{}$\mathrm{Fmmm}$ } & \textcolor{red}{\tiny{}$\mathrm{C2/m}$ (17)}\tabularnewline
\midrule 
{\tiny{}Li} & - & {\tiny{}$\mathrm{Im\overline{3}m}$ (0)} & - & - & {\tiny{}$\mathrm{Im\overline{3}m}$ (0)}\tabularnewline
\bottomrule
\end{tabular}
\par\end{raggedright}{\tiny \par}
\begin{centering}
{\tiny{}$^{*}$The stable compounds of Li-Sn at ambient pressure (\textasciitilde{}
0 GPa) and 0 K temperature are represented by a star ($*$). }
\par\end{centering}{\tiny \par}
{\tiny{}$^{\#}$ Provided format of CIF file is wrong. }{\tiny \par}

{\tiny{}$^{\#\#}$ Fitness of Li-Sn compounds (vertical distance from
the convex hull tie line) given in Table 1 in the results Mayo et
al.\cite{Martin@2017} are put in parentheses. }{\tiny \par}
\end{table}

\begin{figure}[t]
\begin{centering}
\includegraphics[scale=0.45]{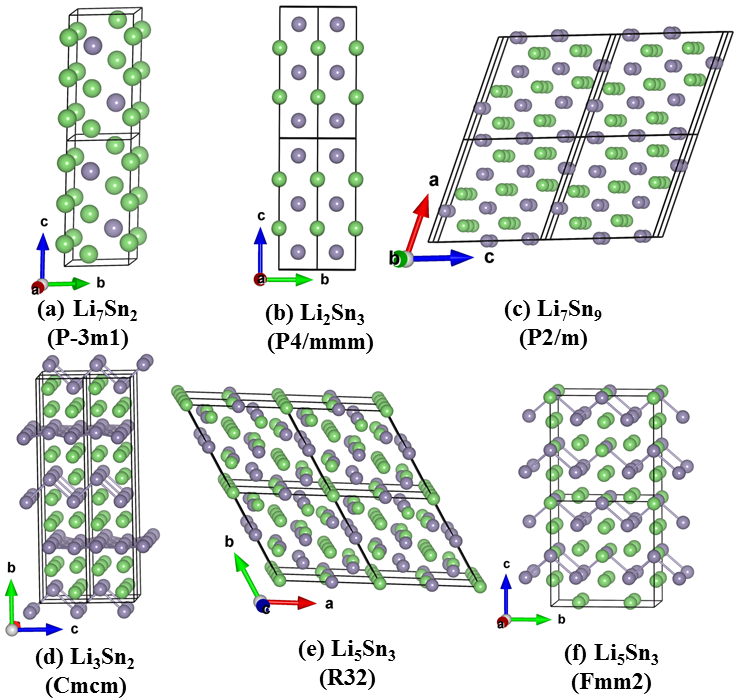}
\par\end{centering}
\centering{}\caption{Lowest energy structures of meta-stable Li-Sn compounds: (a) $\mathrm{Li_{7}Sn_{2}}$
($\mathrm{P\overline{3}m1}$), (b) $\mathrm{Li_{2}Sn_{3}}$ ($\mathrm{P4/mmm}$),
(c) $\mathrm{Li_{7}Sn_{9}}$ ($\mathrm{P2/m}$), (d) $\mathrm{Li_{3}Sn_{2}}$
($\mathrm{Cmcm}$), (e) $\mathrm{Li_{5}Sn_{3}}$ ($\mathrm{R32}$),
and (f) $\mathrm{Li_{5}Sn_{3}}$ ($\mathrm{Fmm2}$). The formation
energy of $\mathrm{Li_{3}Sn_{2}}$ ($\mathrm{Cmcm}$) is found comparable
with the phase $\mathrm{Li_{3}Sn_{2}}$ ($\mathrm{P2_{1}/m}$), as
reported by Mayo et al.\cite{Martin@2017} In case $\mathrm{Li_{5}Sn_{3}}$,
current calculations have found two polymorphs ($\mathrm{R32}$ and
$\mathrm{Fmm2}$) with same formation energy at ambient pressure (\textasciitilde{}
0GPa) and 0 K temperature. In the figure, green and purple spheres
denote the Li and Sn atoms, respectively.\label{structure}}
\end{figure}

\textcolor{black}{Mayo et al. have described the symmetry of $\mathrm{Li_{2}Sn_{3}}$
to be triclinic with space group $\mathrm{P\overline{1}}$ everywhere
in the article. But, in the ``Supporting Information'', they have
provided the CIF file for $\mathrm{P4/mmm}$ phase. Thus, the structural
information described in the manuscript mislead the exact crystallographic
symmetry. Present calculations, however, predict the correct ground
state structure of $\mathrm{Li_{2}Sn_{3}}$ to be $\mathrm{P4/mmm}$,
which is shown in Figure \ref{structure} (b). The atomic arrangement
of $\mathrm{Li_{2}Sn_{3}}$ - $\mathrm{P4/mmm}$ is very similar to
$\mathrm{Li_{1}Sn_{2}}$ -$\mathrm{P4/mmm}$. The only difference
is, instead of double layers of four-membered rings of Sn, single
layers of four-membered rings of Sn can also be found in the $\mathrm{Li_{2}Sn_{3}}$
- $\mathrm{P4/mmm}$ structure in an alternative sequence, as shown
in Figure \ref{structure} (b). The absence of imaginary phonon frequency
in the whole Brillouin zone as shown in Figure S1 (a) in SI represents
the dynamical stability of $\mathrm{Li_{2}Sn_{3}}$ -$\mathrm{P4/mmm}$
at ambient pressure condition.}

\textcolor{black}{In their article, Mayo et al. have described $\mathrm{Li_{7}Sn_{9}}$
to exhibit $\mathrm{P4_{2}/n}$ symmetry, with ground state formation
enthalpy lying about 19 meV/atom above the convex hull tie-line. However,
in the CIF file they have provided the structure for $\mathrm{C222}$
symmetry. On the contrary, current calculations discovered a new structure
having symmetry $\mathrm{P2/m}$. This structure has been found to
be energetically much stable than $\mathrm{C222}$ phase (with 159
meV/atom difference in formation energy). Additionally, this structure
has been found just 7 meV/atom above the convex hull tie line (Figure
\ref{1atm_chull}). In this structure, 1D infinite atomic chains of
pure Li and Sn, can be found to run along {[}010{]} direction (Figure
\ref{structure} (c)). Phonon dispersion curve (See Figure S1 (b)
in SI) represents $\mathrm{Li_{7}Sn_{9}}$ - $\mathrm{P2/m}$ is also
dynamically stable.}

\textcolor{black}{In case of $\mathrm{Li_{3}Sn_{2}}$, $\mathrm{P2_{1}/m}$
symmetry is predicted by Mayo et al.\cite{Martin@2017} While, our
calculations predict the ground state structure of $\mathrm{Li_{3}Sn_{2}}$
to exhibit $\mathrm{Cmcm}$ phase. It is found that in a tight relaxation
of Hellmann-Feynman forces, $\mathrm{P2_{1}/m}$ symmetry transforms
into $\mathrm{Cmcm}$ symmetry. In $\mathrm{Li_{3}Sn_{2}}$-$\mathrm{Cmcm}$
structure, Li atoms can be seen to intercalate between 1D zigzag chains
and 2D buckled layer of Sn ions (Figure \ref{structure} (d)). Absence
of imaginary frequency in phonon dispersion curve revels the dynamical
stability of $\mathrm{Li_{3}Sn_{2}}$-$\mathrm{Cmcm}$ phase, shown
in Figure S1 (c) in SI.}

\textcolor{black}{The described symmetry of $\mathrm{Li_{5}Sn_{3}}$
highly misleads the actual crystallographic informations of the structure.
In the ``ABSTRACT'' and ``Table 1'' of their article, Mayo et
al. have described the symmetry to be $\mathrm{I\mathrm{m\overline{3}m}}$,
while in ``RESULTS'' section, they discussed the $\mathrm{P\overline{4}3m}$
symmetry of $\mathrm{Li_{5}Sn_{3}}$, and have provided its CIF file.
Interestingly, on relaxing the $\mathrm{Li_{5}Sn_{3}}$-$\mathrm{P\overline{4}3m}$
structure provided by Mayo et al.,\cite{Martin@2017} it has been
found energetically unstable (Figure \ref{1atm_chull}). On the other
hand, current calculations predict a new phase having space group
$\mathrm{R32}$ within the rhombohedral symmetry. The formation enthalpy
of this phase is found to be lower than both, $\mathrm{I\mathrm{m\overline{3}m}}$
and $\mathrm{P\overline{4}3m}$ phase of $\mathrm{Li_{5}Sn_{3}}$
(Table \ref{convex-hull-table}). Other than this, one more polymorph
of $\mathrm{Li_{5}Sn_{3}}$ having symmetry $\mathrm{Fmm2}$ is found,
whose formation enthalpy is as comparable as $\mathrm{R32}$ phase
of $\mathrm{Li_{5}Sn_{3}}$. In $\mathrm{R32}$ phase, two different
type of 1D atomic chains, one with pure Li another having an alternative
sequence of Li and Sn atoms, can be seen to run along $c$-axis (shown
in Figure \ref{structure} (e)). While, the crystal structure of $\mathrm{Fmm2}$
is very much unusual, where bent trimers of pure Sn stacked one after
another along crystallographic $a$-axis, are surrounded by Li atoms
(Figure \ref{structure} (f)). The phonon dispersion curves of both
of these newly predicted phases ($\mathrm{R32}$ and $\mathrm{Fmm2}$)
are provided in SI (Figure S1 (d) \& (e)). It should also be noted
here that few years back, Courtney et al.\cite{Courtney@1998} revealed
through experimental study that all known phases of Li-Sn do not always
form in the electrochemical cell of Sn anode, operating at room temperature.
Their investigation indicates that for $\mathrm{x}>2.5$ in $\mathrm{Li_{x}Sn}$,
one of the unknown phases of Li-Sn adopts cubic like structure. In
predicting new stable and metastable Li-Sn compounds, Mayo et al.
therefore suggest that this cubic like arrangement may arise due the
formation of $\mathrm{Li_{5}Sn_{3}}$, as they have showed that the
ground state structure of $\mathrm{Li_{5}Sn_{3}}$ would be $\mathrm{Im\overline{3}m}$.
But, in this comment, it has been showed that the ground state structure
of $\mathrm{Li_{5}Sn_{3}}$ is rhombohedral and not cubic. Moreover,
in our recent work,\cite{Sen@2017} we have showed that at ambient
pressure and room temperature, $\mathrm{Li_{3}Sn_{1}}$ forms a cubic
phase ($\mathrm{P2/m}$ $\stackrel{311\mathrm{K}}{\longrightarrow}$$\mathrm{Fm\overline{3}m}$),
which is also consistent with results of Thackeray et al.\cite{Thackeray@2003}
Therefore, it may be concluded in this comment that the formation
of cubic like arrangements in electrochemical cell may appears due
to the formation of $\mathrm{Li_{3}Sn_{1}}$-$\mathrm{Fm\overline{3}m}$
phase, where Li atoms form a bcc-like sub-lattice.}

\textcolor{black}{For $\mathrm{Li_{3}Sn_{1}}$, we again notice a
huge inconsistency in the results. Mayo et al. described the symmetry
to be $\mathrm{P3_{2}}$ in the ``RESULTS'' section and Table 1
of their manuscript,\cite{Martin@2017} while provided the structure
for $\mathrm{P3m1}$ symmetry in the ``Supporting Information''
and haven't discussed its structure at all. This structure, however,
on relaxation is found unstable. In contrast, our search predicts
$\mathrm{P2/m}$ symmetry for $\mathrm{Li_{3}Sn_{1}}$ to be much
more energetically stable than $\mathrm{P3m1}$ phase. The $\mathrm{Li_{3}Sn_{1}}$-
$\mathrm{P2/m}$ structure is found to lie just 7 meV/atom above our
convex hull tie line (Figure\ref{1atm_chull} and Table \ref{convex-hull-table}).
The structure of $\mathrm{Li_{3}Sn_{1}}$-$\mathrm{P2/m}$ has been
described in our recent work.\cite{Sen@2017} A one-to-one comparison
with $\mathrm{Li_{3}Sn_{1}}$-$\mathrm{P3_{2}}$ could not be made
on the basis of formation enthalpy because of unavailibity of its
crystal structure. However, we are confident that the lowest energy
structure will correspond only to the $\mathrm{Li_{3}Sn_{1}}$- $\mathrm{P2/m}$
symmetry,}

\textcolor{black}{In case of $\mathrm{Li_{7}Sn_{2}}$, experimentally
known phase is $\mathrm{\mathrm{Cmmm}}$, while we found $\mathrm{P\overline{3}m1}$
phase (Figure \ref{structure} (a)) to be more stable than $\mathrm{\mathrm{Cmmm}}$.\cite{Sen@2017}
Though the difference in the enthalpy for both phases is just $6$
meV/atom, which is also in agreement with results of Geneser et al.\cite{Genser_2001}
The structure is also shown to be dynamically stable in our recent
article.\cite{Sen@2017} On the contrary, Mayo et al.\cite{Martin@2017}
predicted $\mathrm{Li_{7}Sn_{2}}$ to be stable under $\mathrm{P\overline{1}}$
symmetry, which they also claim to be 6 meV/atom lower in formation
energy than the experimentally reported $\mathrm{\mathrm{Cmmm}}$
phase. However, it has been noticed that the format of the CIF file
is completely wrong (it is in ``SHELX'' format). On analyzing the
given format, it is found that the structure given in their CIF file
is actually representing the $\mathrm{P\overline{3}m1}$ phase. Therefore,
the discussion about $\mathrm{Li_{7}Sn_{2}}$ in the Ref. {[}1{]}
remains irrelevant, as it is providing a misleading crystallographic
information. }

\textcolor{black}{In case of $\mathrm{Li_{4}Sn_{1}}$, Mayo et al.
have mentioned the $\mathrm{P2_{1}}$ symmetry in the Abstract and
Table1 but provided the CIF file for the $\mathrm{R\overline{3}m}$
phase, which is the lowest energy structure. Thus, their paper provide
a wrong and contradicting information about the symmetry in the text. }

\textcolor{black}{The information provided for $\mathrm{Li_{5}Sn_{1}}$,
is again highly contradictory. In ``Abstract'' Mayo et al. have
described the symmetry to be $\mathrm{P6/mmm}$ (19 meV/atom above
the convext hull tie line) and have provided its CIF file as well.
However, in ``Table 1'' of their article they presented the structure
symmetry to be $\mathrm{Pmma}$. In contract, we found the symmetry
of structures to be $\mathrm{C2/m}$, which is found 11 meV/atom above
from the convex-hull tie line. This clearly indicates $\mathrm{C2/m}$
phase to be energetically favorable, as compared to $\mathrm{P6/mmm}$
phase. }

\textcolor{black}{In case of $\mathrm{Li_{7}Sn_{1}}$, Mayo et al.
described the symmetry to be $\mathrm{C2}$, while in ``Table 1'',
as well as in the CIF file, they represented the $\mathrm{Fmmm}$
phase for this compound. On the other hand, our calculation predicted
$\mathrm{C2/m}$ phase for $\mathrm{Li_{7}Sn_{1}}$.\cite{Sen@2017}
The formation enthalpy of our predicted structure ($\mathrm{C2/m}$)
is found to be as comparable as the $\mathrm{Fmmm}$ structure, reported
by Mayo et al.\cite{Martin@2017} The $\mathrm{C2/m}$ phase is also
shown to be dynamically stable in our recent study.\cite{Sen@2017}}

\textcolor{black}{It is also noteworthy to mention here that Mayo
et al. described $\mathrm{Li_{15}Sn_{4}}$ as a known phase in the
Li-Sn binary phase diagram. But, to the best of our knowledge, we
did not find any such compound for Li-Sn from our calculations. The
same has not even be available in the ICSD database, as well. However,
in cross verification of the reference $\mathrm{Li_{15}Sn_{4}}$,\cite{Zintl@1936}
given by Mayo et al., it has been observed that the reference is actually
describing the structure of $\mathrm{Na_{15}Sn_{4}}$, rather than
$\mathrm{Li_{15}Sn_{4}}$.}

\section*{Conclusions }

\textcolor{black}{In summary, crystal structure and symmetry of several
Li-Sn compounds reported by Mayo et al.\cite{Martin@2017} are not
only contradicting within the article (e.g., $\mathrm{Li_{2}Sn_{3}}$,
$\mathrm{Li_{7}Sn_{9}}$, $\mathrm{Li_{5}Sn_{3}}$, $\mathrm{Li_{3}Sn_{1}}$,
$\mathrm{Li_{7}Sn_{2}}$, $\mathrm{Li_{4}Sn_{1}}$, $\mathrm{Li_{5}Sn_{1}}$,
and $\mathrm{Li_{7}Sn_{1}}$,) but in many cases (e.g., $\mathrm{Li_{7}Sn_{9}}$,
$\mathrm{Li_{5}Sn_{3}}$, $\mathrm{Li_{3}Sn_{1}}$ and $\mathrm{Li_{5}Sn_{1}}$)
they are wrong as well. Thus, their claim of understanding the lithiation
mechanism in Sn anode is incorrect and the conclusions are questionable.
We hope that this comment will be helpful for the community to identify
the correct ground state structure and phase of various Li-Sn compounds
that may appear during lithiation-delithiation of Sn anode.}
\begin{acknowledgement}
P. J. acknowledges the support provided by Grant No. SR/FTP/PS-052/2012
from Department of Science and Technology (DST), Government of India.
R.S. wants to thank Mr. Dwaipayan Chakroborty for critical reading
of the manuscript. The high performance computing facility and workstations
available at the School of Natural Sciences, Shiv Nadar University,
were used to perform all calculations.
\end{acknowledgement}
\begin{suppinfo}
\textcolor{black}{Methodology and Computational Details, Phonon dispersion
curves of Li-Sn compounds (pdf), C}IF files of newly predicted structures
(ZIP)
\end{suppinfo}
\bibliography{refs}

\end{document}